# Surface fractal dimension and its theoretical relationship with adsorbed water content

Behzad Ghanbarian

Department of Geology, Kansas State University, Manhattan 66506 KS, USA

Email: ghanbarian@ksu.edu

**Abstract**

Surface fractal dimension $D_s$ is a quantity describing the roughness of pore-solid interface where all interactions between solid matrix and fluid in the pore space occur. $D_s$ also quantifies surface area; the higher the surface fractal dimension the greater the surface area. Therefore, at some high enough tension head, where a thin layer of water covers the pore-solid interface, one should expect adsorbed water content to be related to $D_s$ in water-wet porous media. In this technical note, we develop a theoretical relationship between the surface fractal dimension, $D_s$, and the adsorbed water content, $\theta_{ads}$, using concepts from van der Waals and electrostatic forces. The proposed model sheds light on constant coefficients of logarithmic equations found empirically between $D_s$ and water contents retained at 1500 and 10000 kPa tension heads. Results also show that our theoretical model estimates $D_s$ from first physical principles for 164 soil samples accurately.

## 1. Introduction

Surface fractal dimension $D_s$ is a quantity describing the roughness of pore-solid interface where all interactions between fluid in the pore space and solid matrix occur. Fluid in the pore space generally exists in three main forms under variably-saturated and water-wet conditions: (1) within the pore space among solid grains, (2) as pendular rings around the contact point of two grains, and (3) in the form of thin films covering the pore-solid interface (see Fig. 1), as conjectured by Buckinhgam (1907). Accordingly, Waldron et al. (1961), Or and Tuller (1999), and many others showed that capillary-based models of porous media which merely incorporate the effect of capillarity would underestimate water content and flow at high tension heads due to neglecting contributions from adsorbed water films. This means understanding film flow, which depends on physicochemical properties of porous media (Tokunaga, 2011), and its mechanism is necessary to study two-phase flow and transport in water-wet porous media, particularly at low water saturations.

In the past two decades, concepts from adsorbed water films have been widely used to investigate different phenomena related to fluid flow and transport in variably-saturated porous materials (see e.g., Nitao and Bear, 1996; Long and Or, 2005; Kim et al., 2012; Heath et al., 2014; Tokunaga et al., 2017). For example, Tokunaga and Wan (1997) demonstrated experimentally that film flow could effectively contribute to total flow in unsaturated fractures for tension heads ranged between 0 and 0.25 kPa. Those authors used a simple model, average velocity as a function of surface film thickness squared proposed by Bird et al. (1960), and found film thicknesses between 2 to 70 μm. This range, however, seems to be much greater than recent results. For instance, Kibbey (2013) used stereoscopic scanning electron microscopy to determine surface-associated water thickness for tension heads ranged from 1 to 10 kPa. His results indicated that water layer thickness varied from about 0.26 to 1.5 μm; see Table 1 in Kibbey (2013). Since the thicknesses of the surface-associated water were found to be orders of magnitude greater than those which might be expected at the same tension heads based on calculations of adsorbed film thickness, Kibbey (2013) concluded that almost all of the surface-associated water was capillary held.

Both pore-solid interface roughness and surface area are characterized by surface fractal dimension $D_s$ such that the higher the surface fractal dimension, the rougher the pore-solid interface, and thus the greater the surface area. Since at high enough tension heads a thin layer of water would cover the pore-solid interface in water-wet porous media, one should expect adsorbed water content to be related to $D_s$. The main objective of this technical note is to develop a theoretical relationship between the surface fractal dimension $D_s$ and the adsorbed water content $\theta_{ads}$ using concepts from van der Waals and electrostatic forces. Interestingly, our proposed model is similar in form to the empirical relationships experimentally found between $D_s$ and water contents retained at 1500 and 10000 kPa tension heads ($\theta_{1500}$ and $\theta_{10000}$, respectively).

## 2. Theory

Thin liquid films adsorbed on the pore-solid interface occur due to several types of surface forces: (1) van der Waals, responsible for molecular interactions, (2) electrostatic, responsible for diffuse double layers, and (3) hydration forces, short-range repulsive forces acting between polar surfaces (Tuller et al., 1999; Saramago, 2010). In this section, we first invoke concepts from adsorbed water films to link disjoining tension to film thickness. For the sake of simplicity, we only consider the effects of van der Waals and electrostatic forces. We then apply the power-law water retention curve model of de Gennes (1985b) to relate tension head and accordingly surface fractal dimension to adsorbed water content.

## 2.1. van der Waals films

Following de Gennes (1985a), Israelachvili (1992) and Iwamatsu and Horii (1996), the relation between disjoining tension, $h_v$ (Pa), and film thickness due to van der Waals forces, $e_v$ (m), is given by

$$h_v = \frac{-A}{6\pi e_v^3} \tag{1}$$

where $A$ is the Hamaker constant (J), which is extremely difficult to measure directly (Baveye, 2012). Note that Eq. (1) is valid for planar surfaces, non-polar fluids, and when interactions are dominantly van der Waals. Although Tokunaga (2011) modified Eq. (1) to include additional compressive capillary pressure due to the curvature of air-water interfaces coating ideal spherical solid particles, he indicated that its effect is small and negligible.

Tuller and Or (2005) invoked concepts from van der Waals forces, estimated soil specific surface area from water retention data at low water content values, and found good agreement with measurements. In that study, Tuller and Or (2005) used an effective value of -6 × 10-20 J for the Hamaker constant $A$. Using soil water retention data corresponding to tensions greater than 10000 kPa and the Tuller and Or (2005) approach, Resurreccion et al. (2011) calculated the Hamaker constant value for 41 Danish soils and found -513 × $10_{-20} \leq A \leq -1 \times 10_{-20}$ J. However, for most low organic soils the Hamaker constant varied between $-10 \times 10_{-20}$ and $-1 \times 10_{-20}$ J, in accord with the results of Tuller and Or (2005) who stated that the Hamaker constant value of $-6 \times 10_{-20}$ J should be seen as an effective value that lumps effects of heterogeneous surface properties, geometry, and van der Waals interactions. Nonetheless, the values of $|A|$ reported by Tuller and Or (2005) and Resurreccion et al. (2011) are greater than those reported in the literature for minerals (see Table 1 of Tokunaga, 2011). This might be due to additional corporations e.g., capillarity, electric double-layer effects, etc. into the soil water retention curve (Tokunaga, 2011).

## 2.2. Electrostatic films

A simple electric double-layer model was proposed by Langmuir (1938) to describe adsorbed film thicknesses for dilute aqueous solutions in equilibrium on a solid of high surface electrical potential. The Langmuir model is

$$h_e = \frac{\varepsilon_0 \varepsilon_r}{2} \left( \frac{\pi k_B T}{E_c z} \right)^2 \frac{1}{e_e^2} \tag{2}$$

where $h_e$ and $e_e$ are the disjoining tension (Pa) and film thickness (m) due to electrostatic forces, respectively, $\varepsilon_0$ is the vacuum permittivity ($= 8.85 \times 10_{-12}$ C$_2$ J$_{-1}$ m$_{-1}$), $\varepsilon_r$ is the dielectric constant of water ($= 78.54$ at 25℃), $E_c$ is the electric charge ($1.602 \times 10_{-19}$ C), $k_B$ is the Boltzmann constant ($1.381 \times 10_{-23}$ J K$_{-1}$), $T$ is the Kelvin temperature, and $z$ is the ion valence. Although experimental evidence indicated that the Langmuir film model starts to deviate from measurements when thickness is less than about 30 nm, it still provides reasonable results (Israelachvili, 1992). Note that Eq. (2) was obtained by assuming that electrostatic potentials are large throughout the film (Tokunaga, 2009). If one includes a capillary scaling term acting on the convex film, the Langmuir model (Eq. 2) can be modified. However, it was shown that this contribution is small and approximations for flat surfaces are enough and suitable (Tokunaga, 2011).

## 2.3. de Gennes (1985b) water retention curve model

In the literature, there exist numerous water retention curve models (Ghanbarian-Alavijeh et al., 2011). Although some of them are empirical, some others are on the basis of hierarchical and self-similar properties of porous media, a power-law pore-size distribution, and a bundle of non-interconnected cylindrical pores. The pioneer water retention curve model of de Gennes (1985b) relates tension head, $h$, to water content, $\theta$, in the following power-law form

$$h = h_d \left( \frac{\theta}{\theta_s} \right)^{D_s - 3} \tag{3}$$

where $D_s$ is the surface fractal dimension ranging between 2 and 3 in three dimensions, $h_d$ is the displacement tension head (Pa), and $\theta_s$ is the saturated water content. Equation (3) was later derived by Tyler and Wheatcraft (1990), Perrier et al. (1996), among others, using different terminologies. For a recent comprehensive review of fractal water retention curve models see Ghanbarian and Millán (2017). One should note that Millán and Gonzales-Posada (2005) as well as Ojeda et al. (2006) demonstrated that Eq. (3) is not only valid at high and intermediate water contents (or low and intermediate tension heads) but also at low water contents (or high tensions).

**2.4. Theoretical relationship between surface fractal dimension and adsorbed water content**

In this section, we invoke the power-law water retention curve model of de Gennes (1985b), Eq. (3), to relate the tension head to the water content and accordingly $D_s$ to the adsorbed water content $\theta_{ads}$. The latter corresponds to $h_{ads}$, some high enough tension head where water mainly exists as adsorbed thin films in the medium. One should expect $h_{ads}$ for coarse-textured soils to be less than that for fine-textured soils.

Individual contributions of capillarity, van der Waals, and electrostatic forces to the tension head are seldom differentiated in the determination of the soil water retention curve (Tuller and Or, 2005). For the sake of simplicity, we assume that the adsorbed water content $\theta_{ads}$ is mainly controlled by van der Waals and electrostatics forces. Consequently, as a *first-order approximation*, we set $h_{ads} \approx h_v + h_e$, meaning that van der Waals and electrostatic forces are dominant over capillarity forces. Thus, combining Eqs. (1) and (2) with the natural logarithmic form of Eq. (3) yields

$$D_s = \frac{1}{C}\ln(\theta_{ads}) + \left(3 - \frac{\ln(\theta_s)}{C}\right) \tag{4}$$

in which

$$C = ln\left[\frac{-A}{6\pi h_d e_v^3} + \frac{\varepsilon_0 \varepsilon_r}{2h_d e_e^2}\left(\frac{\pi k_B T}{E_c z}\right)^2\right]$$

where $e_v$ (m) and $e_e$ (m) are the adsorbed film thicknesses due to van der Waals and electrostatic forces, respectively, at $h = h_{ads}$.

Interestingly, Eq. (4) is similar in form to the following logarithmic equation found empirically by Ghanbarian-Alavijeh and Millán (2009)

$$D_s = 0.1055\ln(\theta_{1500}) + 3.0298 \tag{5}$$

where $\theta_{1500}$ is the water content at 1500 kPa tension head. Ghanbarian-Alavijeh and Millán (2009) anticipated the contribution of van der Waals forces at high tension heads. However, those authors never applied its concepts to theoretically relate $D_s$ to $\theta_{1500}$ to justify the physical meaning of the constant coefficients in their empirical relationship, Eq. (5). Similarities between Eq. (4), derived theoretically using concepts from adsorbed liquid films, and Eq. (5), proposed experimentally using 172 soil samples, shed light on the two empirical coefficients 0.1055 and 3.0298 in Eq. (5), as we will discuss in the Results and Discussion section.

## 3. Experimental data

In this study, we use five databases, namely Puckett et al. (1985), UNSODA (Leij et al., 1996), GRIZZLY (Haverkamp et al., 1997), Huang et al. (2006), and Fooladmand (2007) including 172 samples. The salient characteristics of the soil samples in each database are presented Table 1. Most $D_s$ values reported in Table 1 range from 2.5 to near 3. However, for eight sandy soil samples from the GRIZZLEY database we found $D_s < 2$, which are not theoretically supported. We accordingly removed them from further analysis. One should note that although $D_s < 2$ is not supported in three dimensions, Ghanbarian-Alavijeh and Hunt (2012) found that negative pore space fractal dimension values are theoretically possible (see also Mandelbrot, 1990).

Or and Tuller (2000) as well as Tuller and Or (2005) argued that for $h \geq 10000$ kPa the capillarity effect would be negligible. Therefore, for practical purposes, we set $h_{ads} = 10000$ kPa. For soil samples for which water content at $h = 33$, 1500 and 10000 kPa was not measured, $\theta_{33}$, $\theta_{1500}$, and $\theta_{10000}$ were determined from Eq. (3) using the measured value of $\theta_s$ and optimized values of $D_s$ and $h_d$. The interested reader is referred to the original published papers and the Ghanbarian-Alavijeh and Millán (2009) article for further information. Table 2 summarizes the values of parameters in Eq. (4) required to link the surface fractal dimension $D_s$ to the adsorbed water content $\theta_{ads}$.

## 4. Results and Discussion

In this section, we first present surface fractal dimension, determined experimentally by fitting Eq. (3) to the measured water retention curves of 164 soil samples summarized in Table 1, versus water contents retained at various tension heads. We then investigate the sensitivity of the theoretical $D_s$-$\theta_{ads}$ relationship, Eq. (4), to input parameters such as Hamaker constant, displacement tension head, film thicknesses due to van der Waals and electrostatic forces, and saturated water content. Using Eq. (4), we next calculate $D_s$ from measured $\theta_s$ and $h_d$ values, water content at 10000 kPa ($\theta_{ads} = \theta_{10000}$), as well as approximated $e_v = e_e = 1$ nm and $A = -6 \times 10_{-20}$ J values.

### 4.1. $D_s$ versus water content at various tension heads

Figure 2 shows surface fractal dimension $D_s$, calculated by directly fitting Eq. (3) to the measured water retention curves, as a function of saturated water content $\theta_s$, and water contents at 33, 1500, and 10000 kPa tension heads. In addition to 164 samples (shown by open blue circles in Fig. 2), for the sake of comparison we present the data of Rawls et al. (1982) who reported the values of $D_s$ (= $3 - \lambda$; $\lambda$ is the pore-size distribution index), $h_d$, $\theta_s$, $\theta_{33}$, and $\theta_{1500}$, averaged over several hundred samples, for 11 soil texture classes (see

Table 1 of Rawls et al., 1982). In Fig. 2, solid black circles and error bars represent data from Rawls et al. (1982) and one standard deviation about the mean, respectively.

Results indicate that $D_s$ is weakly correlated to $\theta_s$ (see Fig. 2a), while as tension head increases the correlation between $D_s$ and water content becomes stronger (compare Fig. 2b with Figs. 2c and 2d). As can be seen, although tension head increased from 1500 kPa in Fig. 2c to 10000 kPa in Fig. 2d, i.e. near one order of magnitude, the correlation coefficient $R_2$ between $D_s$ and water content only increased from 0.92 to 0.95. This means variation in water content due to tension head increase is not significant at the dry end of soil water retention curve. We found that $\theta_{10000} = 0.95\theta_{1500} - 0.027$ with $R_2 = 0.94$ (results not shown). Figure 2d clearly shows that $D_s$ is highly correlated to $\theta_{10000}$ in a logarithmic form. More specifically, we experimentally found that $D_s = 0.111\ln(\theta_{10000}) + 3.09$ with $R_2 = 0.95$, which is similar in form to our theoretical model, Eq. (4).

### 4.2. Sensitivity to the $D_s - \theta_{ads}$ model parameters

Figure 3 shows the sensitivity of the proposed theoretical $D_s$-$\theta_{ads}$ relationship, Eq. (4), to input parameters $A$ (Hamaker constant), $h_d$ (displacement tension head), $e_v$ (film thickness due to van der Waals forces), $e_e$ (film thickness due to electrostatic forces), and $\theta_s$ (saturated water content). In Fig. 3a, we show that for $\theta_s = 0.5$ cm$_3$ cm$_{-3}$, $h_d = 9.8$ kPa, $e_v = e_e = 1$ nm (equivalent to roughly three water layers) as $|A|$ increases from $6 \times 10_{-21}$ to $6 \times 10_{-19}$, $D_s$ value increases as well. Results indicate that the surface fractal dimension is relatively sensitive to the Hamaker constant $A$.

Figure 3b displays the sensitivity of Eq. (4) to the displacement tension head. We set $\theta_s = 0.5$ cm$_3$ cm$_{-3}$, $A = -6 \times 10_{-20}$ J, and $e_v = e_e = 1$ nm and found that as $h_d$ increases from 0.1 to 10 kPa, $D_s$ decreases. However, we did not find any trend between $D_s$ and $h_d$ for 164 soil samples used in this study (results not shown). Generally speaking, one may expect in the absence of soil structure, the finer the soil texture, the larger the displacement tension head, and the higher the surface fractal dimension (see Table 1). We should point out that $D_s$ is not only a function of $h_d$, but a complicated function of several factors (see Eq. (4)).

In Figs. 3c and 3d, we show the sensitivity of the model to parameters $e_v$ and $e_e$ for a soil sample with $\theta_s = 0.5$ cm$_3$ cm$_{-3}$, $A = -6 \times 10_{-20}$ J, and $h_d = 9.8$ kPa. In Fig. 3c $e_e = 1$ nm and $e_v$ varies from 0.1 to 10 nm, while in Fig. 3d $e_v = 1$ nm and $e_e$ changes between 0.1 and 10 nm. Results indicate that as the film thickness decreases, the surface fractal dimension increases, meaning that when the film thickness, either $e_v$ or $e_e$, is underestimated, the surface fractal dimension $D_s$ is overestimated (see Figs. 3c and 3d). This is consistent with the fractal scaling of surface area (SA) of an individual rough grain of an average radius $R$. In such a case, the surface area (SA) depends on both grain average radius $R$ and probing molecule size $\delta$ as follows (Borkovec et al., 1993)

$$SA \sqcup \delta^{2-D_s} R^{D_s} \tag{6}$$

In Eq. (6), $2 \leq D_s < 3$ which results in an inverse proportionality between SA and $\delta$. This means SA increases as the probing molecule size $\delta$ decreases. For grains with smooth surfaces $D_s = 2$, and thus Eq. (6) reduces to $SA \propto R^2$. Given that $2 \leq D_s < 3$, the surface area tends to infinity ($SA \to \infty$) as the probing molecule size approaches zero ($\delta \to 0$), consistent with Mandelbrot's definition of path lengths on geometrical fractals (Mandelbrot, 1983; Wheatcraft and Tyler, 1988).

Figure 3e shows the calculated $D_s$ using Eq. (4) with $A$ = -6 × 10$_{-20}$ J, $h_d$ = 9.8 kPa, $e_v$ = 1 nm, $e_e$ = 1 nm, and three values of $\theta_s$ (i.e., 0.1, 0.3, and 0.5 cm$_3$ cm$_{-3}$). As can be observed, $\theta_s$ has a relatively substantial impact on $D_s$. This figure also displays that the larger the porosity, the smaller the surface fractal dimension. We, however, did not experimentally find a specific trend between $D_s$ and $\theta_s$ for 164 soil samples used in this study (see Fig. 2a). This is because $D_s$ is a complicated function of several factors, one of which is $\theta_s$ (see Eq. (4)).

### 4.3. $D_s$ estimation from measured $\theta_s$ and $h_d$

Recently, Ghanbarian et al. (2016) presented a theoretical scaling of Poiseuille's approximation for flow in pores with rough surfaces and experimentally showed that incorporating the effect of surface fractal dimension could improve flow predictions in porous media. Accordingly, characterizing the surface fractal dimension for a medium has applications. In this section, we address $D_s$ estimation from measured $\theta_s$ and $h_d$ values. More specifically, we discuss that individual contributions of van der Waals and electrostatic forces to the adsorbed water film thickness are not practically differentiable. The thickness of adsorbed water films is controlled by several factors, such as ionic strength of adsorbed water, spatial variability of mineral surface chemistry, and the topographically complex grain surfaces (Kim et al., 2012). Experimental results of Beaglehole et al. (1991), Asay and Kim (2005) and Bohr et al. (2010), cited in Baveye (2012), indicate that the adsorbed water film thickness at high enough tension heads (i.e., near saturation in the vapor phase) tends to be in the order of about 2 nm, in accord with theoretical results of Tokunaga (2011; 2012). We accordingly set $e_v = e_e = 1$ nm ($e_v + e_e = 2$ nm) and calculate surface fractal dimension using Eq. (4) from measured $\theta_s$ and $h_d$ values, water content at $h = 10000$ kPa ($\theta_{ads} \approx \theta_{10000}$), and $A = -6 \times 10^{-20}$ J (Tuller and Or, 2005). The values of $\varepsilon_0$, $\varepsilon_r$, $\rho_w$, $E_c$, $k_B$, $T$, and $z$ are given in Table 2.

Figure 4 shows the calculated $D_s$ values versus the measured values, determined by directly fitting Eq. (3) to the measured water retention for 164 soil samples used in this study. Although $e_v = e_e = 1$ nm (equivalent to roughly three water layers) and $A = -6 \times 10^{-20}$ J are rough approximations, Eq. (4) estimates the surface fractal dimension from the measured $\theta_s$ and $h_d$ values accurately. As can be observed in Fig. 4, the surface fractal dimension estimates are well around the 1:1 line over the entire range of $D_s$. Discrepancies between the calculated and measured surface fractal dimensions might be due to several factors that are discussed in the following.

The values of $e_v$ and $e_e$ are not essentially the same and equal for all 164 experiments summarized in Table 1, but most probably vary from one sample to another. Note that precise estimation of the adsorbed water film thickness requires high-resolution X-ray nano-computed tomography techniques. In addition, the van der Waals and Langmuir models, respectively Eqs. (1) and (2), are valid for planar surfaces. However, the pore-solid interface of porous media (e.g., soils) is not necessarily smooth and planar. This also might affect $D_s$ estimates via Eq. (4) up to some extent. The Hamaker constant $A = -6 \times 10^{-20}$ (J) reported by Tuller and Or (2005) is also an average value which includes effects of several factors, such as heterogeneity of surface properties, geometry, etc. Its value, thus, might change from one soil sample to another, depending on mineralogical characteristics.

Another reason for discrepancy between the calculated and measured surface fractal dimensions is that, in addition to adsorbed thin water films, at 10000 kPa tension head ($\theta_{10000}$) water may exist in nanopores and/or in depressions on rough pore-solid interface of minerals (Kim et al., 2012). The composite effects of adsorption and capillarity in porous media with rough mineral surfaces may cause deviations from Eq. (4). Furthermore, one should expect wetting properties to vary from one soil sample to another. Results of Wenzel (1936) and Hu et al. (2008) indicate that wettability is a strong function of surface roughness. The latter is positively correlated to water film thickness (see e.g., Chiarello et al., 1993; Bohr et al., 2010). This means that adsorbed water film thickness should be a function of wettability and thus vary from one porous medium to another.

## 5. Conclusion

Using concepts from adsorbed water films and surface forces, we developed a theoretical model, Eq. (4), relating the surface fractal dimension $D_s$ to the adsorbed water content $\theta_{ads}$. More specifically, we used the van der Waals model, Eqs. (1), and the Langmuir equation, Eq. (2), to link the disjoining tension to the adsorbed film thickness. The water retention curve model of de Gennes (1985b) was also used to describe tension head as a function of water content. We further assumed that at some high enough tension head (called $h_{ads}$), where water mainly exists as adsorbed thin films in the medium, van der Waals and electrostatic forces are dominant over capillary forces. Consequently, we set $h_{ads} \approx h_v + h_e$ in which $h_v$ and $h_e$ are disjoining tensions due to van der Waals and electrostatic forces. We showed that the proposed $D_s$-$\theta_{ads}$ model, Eq. (4), well described the experimental observations and the relationship between $D_s$ and water content at 10000 kPa tension head for 164 soil samples. Accurate estimation of the surface fractal dimension using concepts from adsorbed water films, however, requires precise determination of the adsorbed water film thickness. This needs modern techniques, such as high-resolution X-ray nano-computed tomography.

**Acknowledgment**

The author is grateful to Kansas State University for supports through faculty startup funds. The data used in this study are available upon request from the author.

**Figures caption**

Fig. 1. Schematic soil water retention curve and various forms of water under variably-saturated and water-wet conditions. Generally speaking, at low tension heads, water exists within the pore space among solid grains. At intermediate tension heads, water exists as pendular rings around the contact point of two grains. At high tension heads, water is available in the form of thin films covering the pore-solid interface. Gray circles represent solid grains, while blue areas denote water.

Fig. 2. Surface fractal dimension, determined from fitting Eq. (3) to the measured water retention curve, versus water content at (a) $h = 0$ kPa (saturated water content), (b) $h = 33$ kPa, (c) $h = 1500$ kPa, and (d) $h = 10000$ kPa for 164 soil samples given in Table 1 (shown by open blue circles). Solid black circles and error bars represent data from Rawls et al. (1982) and one standard deviation about the mean, respectively.

Fig. 3. Surface fractal dimension determined using Eq. (4) as a function of adsorbed water content. (a) $\theta_s = 0.5$ cm$_3$ cm$_{-3}$, $h_d = 9.8$ kPa, $e_v = 1$ nm, and $e_e = 1$ nm. (b) $\theta_s = 0.5$ cm$_3$ cm$_{-3}$, $A = -6 \times 10_{-20}$ J, $e_v = 1$ nm, and $e_e = 1$ nm. (c) $\theta_s = 0.5$ cm$_3$ cm$_{-3}$, $A = -6 \times 10_{-20}$ J, $h_d = 9.8$ kPa, and $e_e = 1$ nm. (d) $\theta_s = 0.5$ cm$_3$ cm$_{-3}$, $A = -6 \times 10_{-20}$ J, $h_d = 9.8$ kPa, $e_v = 1$ nm. (e) $A = -6 \times 10_{-20}$ J, $h_d = 9.8$ kPa, $e_v = 1$ nm, and $e_e = 1$ nm. Note that $e_v$ and $e_e$ correspond to film thicknesses due to van der Waals and electrostatic forces at some high enough tension head $h_{ads}$. The values of $\varepsilon_0$, $\varepsilon_r$, $E_c$, $k_B$, $T$, and $z$ are given in Table 2.

Fig. 4. Surface fractal dimension calculated via Eq. (4) using measured $\theta_s$ and $h_d$ values, water content at $h = 10000$ kPa ($\theta_{10000}$), $e_v = e_e = 1$ nm, and $A = -6 \times 10_{-20}$ J, versus surface fractal dimension determined by directly fitting Eq. (3) to the measured water

retention curves for 164 soil samples reported in Table 1. The values of $\varepsilon_0$, $\varepsilon_r$, $\rho_w$, $E_c$, $k_B$, $T$, and $z$ are given in Table 2. Although $e_v = 1$ nm, $e_e = 1$ nm, and $A = -6 \times 10^{-20}$ J are approximations, Eq. (4) estimates the surface fractal dimension from measured $\theta_s$ and $h_d$ values accurately. The red dashed line represents the 1:1 line.

Table 1. Salient properties of 172 soil samples used in this study.

| Reference | Texture | No. of samples | Clay (%) | | $\theta_{1500}$ (cm$_3$ cm$_{-3}$) | | $D_s$ | | $R_2$ | |
|---|---|---|---|---|---|---|---|---|---|---|
| | | | Min | Max | Min | Max | Min | Max | Min | Max |
| Huang et al. (2006) | Silt loam | 1 | - | 17.6 | - | 0.114 | - | 2.803 | - | 0.963 |
| | Loamy sand | 2 | 3 | 9.2 | 0.023 | 0.032 | 2.497 | 2.563 | 0.942 | 0.954 |
| | Loam | 5 | 12.2 | 16.4 | 0.076 | 0.104 | 2.745 | 2.771 | 0.99 | 0.993 |
| | Clay loam | 1 | - | 33.5 | - | 0.152 | - | 2.789 | - | 0.99 |
| | Clay | 1 | - | 45.2 | - | 0.212 | - | 2.856 | - | 0.981 |
| Fooladmand (2007) | Silty clay loam | 8 | 28 | 39 | 0.116 | 0.224 | 2.835 | 2.891 | 0.98 | 0.999 |
| | Silty clay | 2 | 42 | 46 | 0.227 | 0.23 | 2.907 | 2.917 | 0.982 | 0.997 |
| | Silt loam | 4 | 12 | 27 | 0.147 | 0.244 | 2.818 | 2.876 | 0.991 | 0.998 |
| | Sandy loam | 2 | 7 | 9 | 0.11 | 0.146 | 2.776 | 2.831 | 0.996 | 0.998 |
| | Loamy sand | 3 | 4 | 6 | 0.89 | 0.1 | 2.761 | 2.807 | 0.993 | 0.996 |
| | Loam | 1 | - | 26 | - | 0.142 | - | 2.847 | - | 0.995 |
| UNSODA | Silty clay loam | 4 | 32 | 35.1 | 0.19 | 0.287 | 2.837 | 2.947 | 0.905 | 0.997 |
| | Silty clay | 4 | 40.3 | 43.5 | 0.154 | 0.278 | 2.832 | 2.96 | 0.856 | 0.993 |
| | Silt loam | 7 | 13.6 | 24.7 | 0.078 | 0.201 | 2.744 | 2.907 | 0.957 | 0.999 |
| | Silt | 1 | - | 9.2 | - | 0.08 | - | 2.802 | - | 0.926 |
| | Sandy clay loam | 2 | 26.8 | 28 | 0.178 | 0.206 | 2.909 | 2.946 | 0.964 | 0.977 |
| | Sandy clay | 2 | 40.5 | 41 | 0.271 | 0.273 | 2.922 | 2.965 | - | 0.937 |
| | Sand | 1 | - | 0.7 | - | 0.02 | - | 2.619 | - | 0.963 |
| | Loamy sand | 3 | 7 | 10.5 | 0.037 | 0.051 | 2.596 | 2.76 | 0.985 | 0.999 |
| | Loam | 7 | 17 | 26.2 | 0.148 | 0.294 | 2.861 | 2.92 | 0.948 | 0.997 |
| | Clay loam | 4 | 29.7 | 38.4 | 0.163 | 0.215 | 2.851 | 2.912 | 0.986 | 0.998 |
| | Clay | 6 | 45 | 63 | 0.285 | 0.414 | 2.941 | 2.969 | 0.846 | 0.995 |
| Puckett et al. (1985) | Sandy loam | 9 | 7.8 | 17.8 | 0.095 | 0.219 | 2.746 | 2.91 | 0.933 | 0.988 |
| | Sandy clay loam | 18 | 20.8 | 42.1 | 0.154 | 0.329 | 2.799 | 2.962 | 0.966 | 0.994 |
| | Sandy clay | 2 | 35.2 | 38 | 0.27 | 0.283 | 2.957 | 2.966 | 0.984 | 0.996 |
| | Sand | 2 | 1.4 | 1.8 | 0.054 | 0.058 | 2.569 | 2.594 | 0.936 | 0.964 |
| | Loamy sand | 5 | 2.3 | 10.8 | 0.062 | 0.136 | 2.607 | 2.837 | 0.897 | 0.984 |
| | Loam | 1 | - | 13.1 | - | 0.167 | - | 2.817 | - | 0.968 |
| | Clay loam | 5 | 30.4 | 34.8 | 0.278 | 0.332 | 2.936 | 2.967 | 0.936 | 0.989 |
| GRIZZLY | Clay | 12 | 43.7 | 77.5 | 0.250 | 0.358 | 2.819 | 2.925 | NA | NA |
| | Clay loam | 2 | 27.0 | 33.9 | 0.114 | 0.150 | 2.793 | 2.844 | NA | NA |
| | Loam | 3 | 12.2 | 20.6 | 0.092 | 0.144 | 2.785 | 2.819 | NA | NA |
| | Loamy sand | 5 | 0.0 | 1.7 | 0.008 | 0.118 | 2.477 | 2.808 | NA | NA |
| | Sand | 15 | 0.0 | 0.0 | 0.000 | 0.064 | 0.409* | 2.747 | NA | NA |
| | Sandy loam | 9 | 0.4 | 12.9 | 0.003 | 0.148 | 2.408 | 2.816 | NA | NA |
| | Silt loam | 3 | 0.6 | 23.8 | 0.077 | 0.109 | 2.700 | 2.792 | NA | NA |
| | Silty clay | 8 | 44.2 | 57.4 | 0.206 | 0.390 | 2.810 | 2.919 | NA | NA |
| | Silty clay loam | 2 | 34.4 | 37.9 | 0.179 | 0.364 | 2.846 | 2.920 | NA | NA |

$\theta_{1500}$ is water content at 1500 kPa tension head; $D_s$ is surface fractal dimension derived by fitting Eq. (3) to the measured water retention curve; $R_2$ is correlation coefficient; NA is not available.

* For eight sandy soil samples in the GRIZZLY database the value of $D_s$ was less than 2, the lower theoretical bound of surface fractal dimension in three dimensions. We accordingly removed those samples from further analysis.

Table 2. Input parameters required for the van der Waals and electrostatic models (Eqs. 1 and 2, respectively).

| parameter (units) | $\varepsilon_0$ (C$_2$ J$^{-1}$ m$^{-1}$) | $\varepsilon_r$ | $E_c$ (C) | $z$ | $T$ (K) | $k_B$ (J K$^{-1}$) |
|---|---|---|---|---|---|---|
| value | $8.85 \times 10^{-12}$ | 78.54 | $1.602 \times 10^{-19}$ | 1 | 298 | $1.381 \times 10^{-23}$ |

$\varepsilon_0$ is vacuum permittivity; $\varepsilon_r$ is dielectric constant of water; $E_c$ is electric charge; $k_B$ is Boltzmann constant; $T$ is Kelvin temperature; $z$ is ion valence.

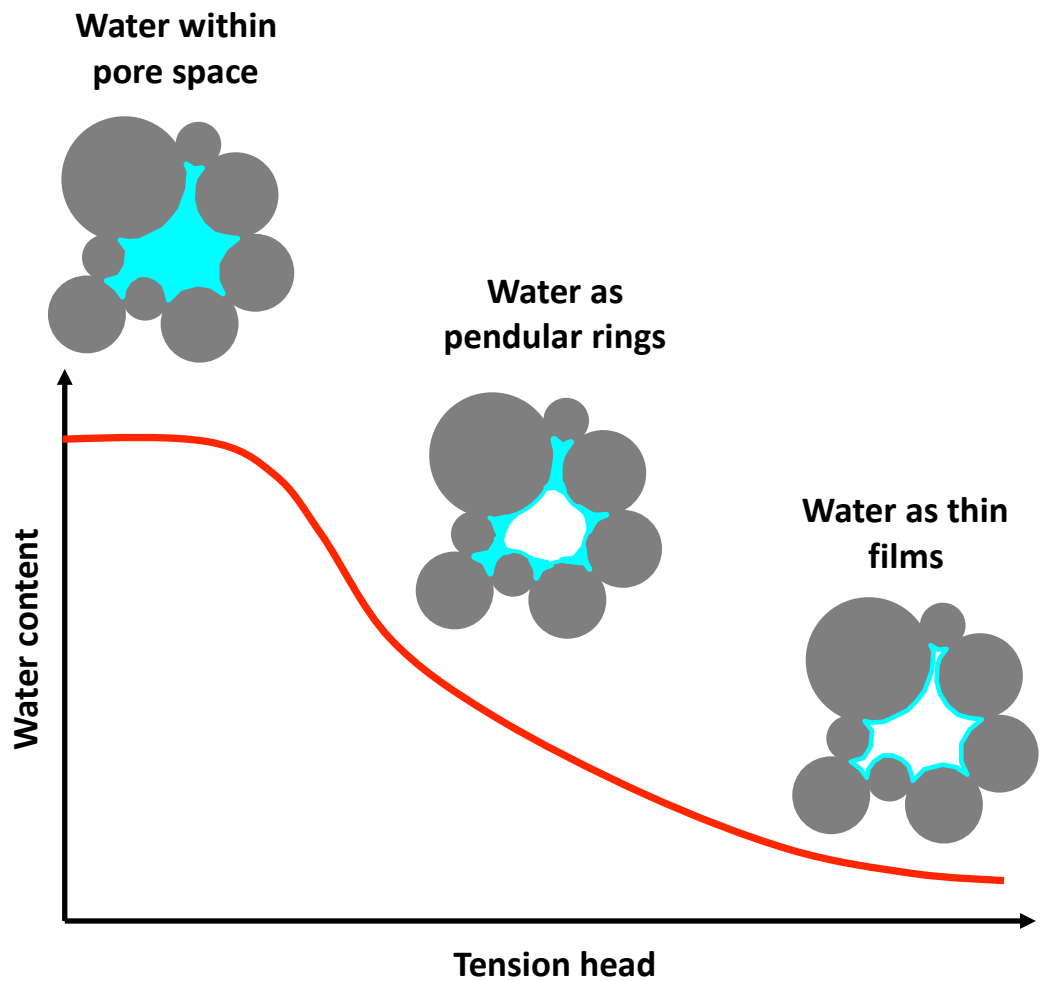

Fig. 1. Schematic soil water retention curve and various forms of water under variably-saturated and water-wet conditions. Generally speaking, at low tension heads, water exists within the pore space among solid grains. At intermediate tension heads, water exists as pendular rings around the contact point of two grains. At high tension heads, water is available in the form of thin films covering the pore-solid interface. Gray circles represent solid grains, while blue areas denote water.

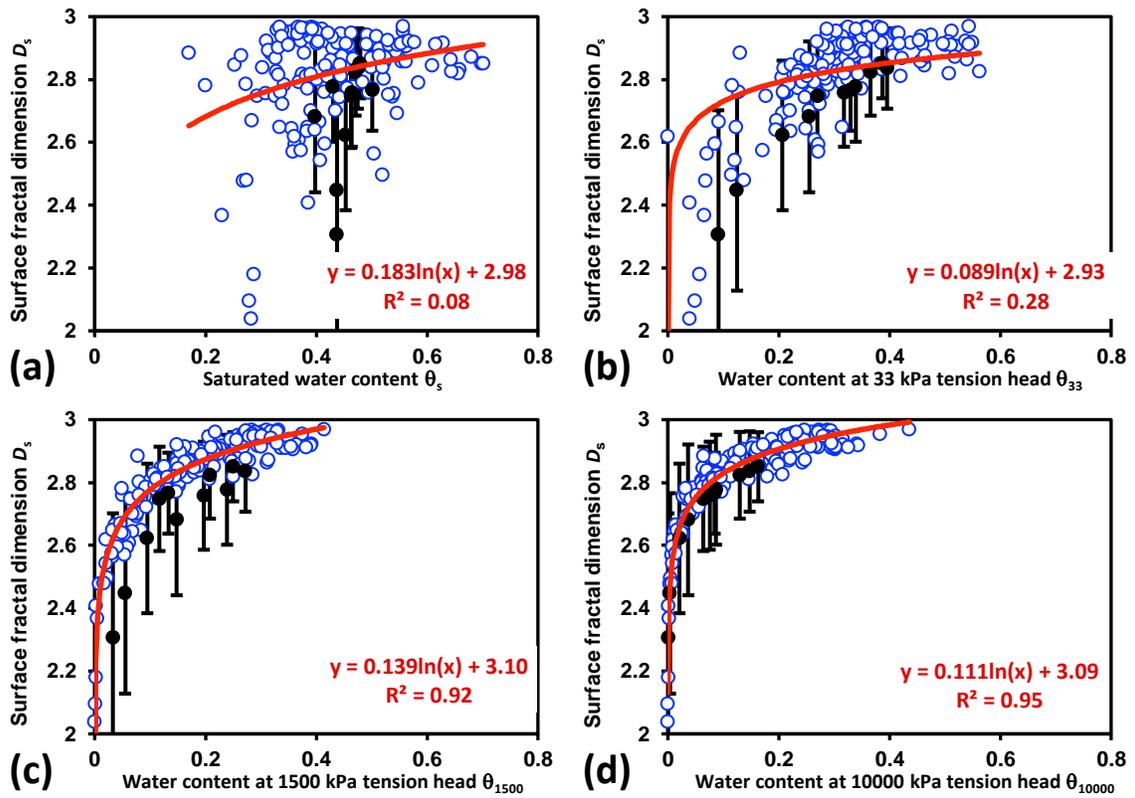

Fig. 2. Surface fractal dimension, determined from fitting Eq. (3) to the measured water retention curve, versus water content at (a) $h = 0$ kPa (saturated water content), (b) $h = 33$ kPa, (c) $h = 1500$ kPa, and (d) $h = 10000$ kPa for 164 soil samples given in Table 1 (shown by open blue circles). Solid black circles and error bars represent data from Rawls et al. (1982) and one standard deviation about the mean, respectively.

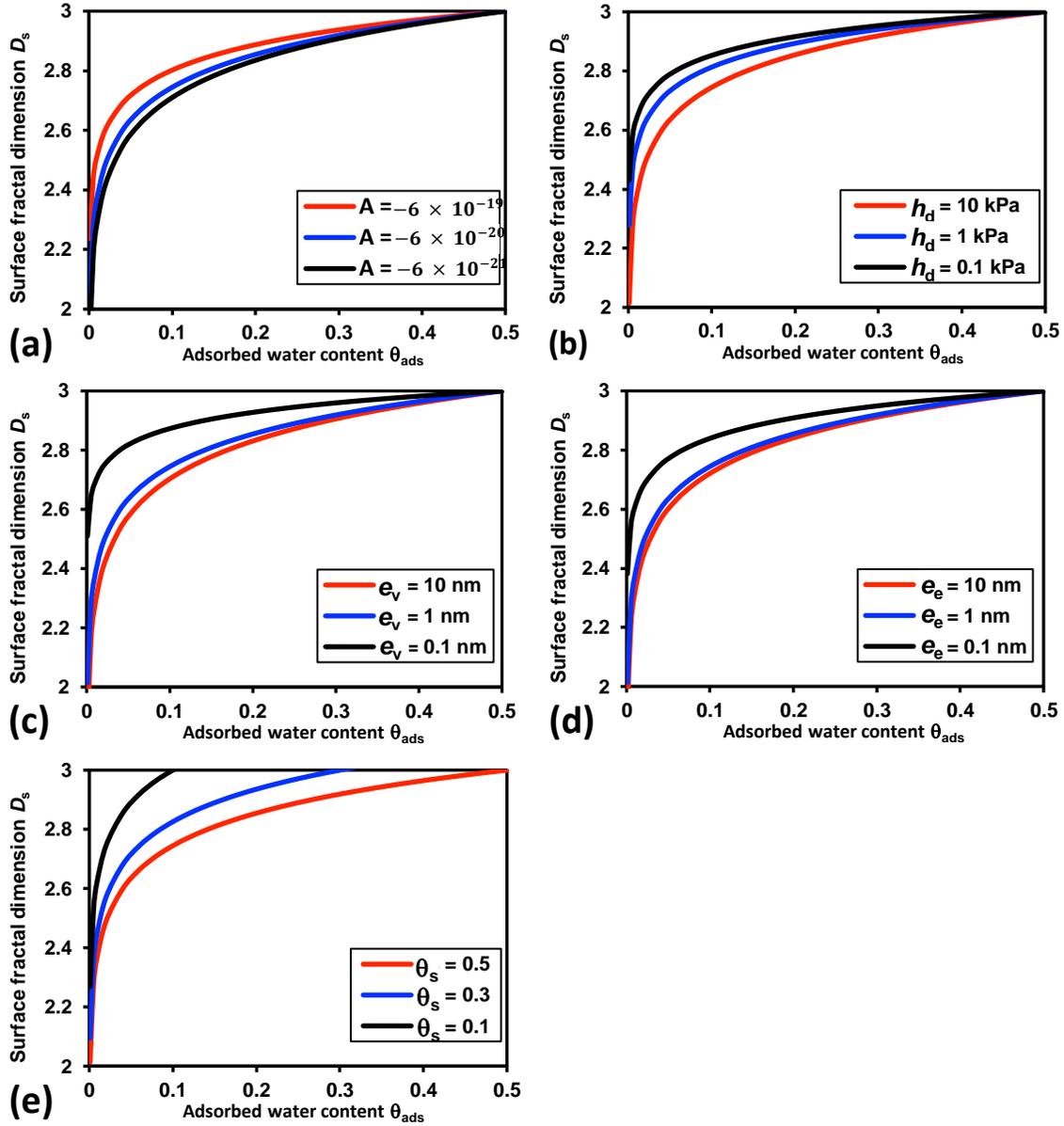

Fig. 3. Surface fractal dimension determined using Eq. (4) as a function of adsorbed water content. (a) $\theta_s$ = 0.5 cm₃ cm-₃, $h_d$ = 9.8 kPa, $e_v$ = 1 nm, and $e_e$ = 1 nm. (b) $\theta_s$ = 0.5 cm₃ cm-₃, $A$ = -6 × 10-₂₀ J, $e_v$ = 1 nm, and $e_e$ = 1 nm. (c) $\theta_s$ = 0.5 cm₃ cm-₃, $A$ = -6 × 10-₂₀ J, $h_d$ = 9.8 kPa, and $e_e$ = 1 nm. (d) $\theta_s$ = 0.5 cm₃ cm-₃, $A$ = -6 × 10-₂₀ J, $h_d$ = 9.8 kPa, $e_v$ = 1 nm. (e) $A$ = -6 × 10-₂₀ J, $h_d$ = 9.8 kPa, $e_v$ = 1 nm, and $e_e$ = 1 nm. Note that $e_v$ and $e_e$ correspond to film thicknesses due to van der Waals and electrostatic forces at some high enough tension head $h_{ads}$. The values of $\varepsilon_0$, $\varepsilon_r$, $E_c$, $k_B$, $T$, and $z$ are given in Table 2.

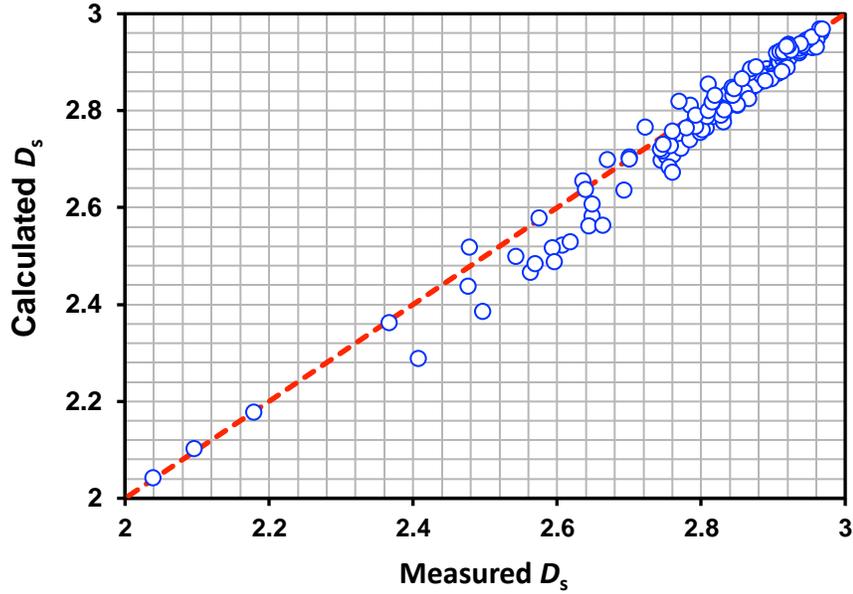

Fig. 4. Surface fractal dimension calculated via Eq. (4) using measured $\theta_s$ and $h_d$ values, water content at $h = 10000$ kPa ($\theta_{10000}$), $e_v = e_e = 1$ nm, and $A = -6 \times 10^{-20}$ J, versus surface fractal dimension determined by directly fitting Eq. (3) to the measured water retention curves for 164 soil samples reported in Table 1. The values of $\varepsilon_0$, $\varepsilon_r$, $\rho_w$, $E_c$, $k_B$, $T$, and $z$ are given in Table 2. Although $e_v = 1$ nm, $e_e = 1$ nm, and $A = -6 \times 10^{-20}$ J are approximations, Eq. (4) estimates the surface fractal dimension from measured $\theta_s$ and $h_d$ values accurately. The red dashed line represents the 1:1 line.